# Dynamics of Photo-excited Hot Carriers in Hydrogenated Amorphous Silicon Imaged by 4D Electron Microscopy

Bolin Liao[a,b], Ebrahim Najafi[a], Heng Li[a], Austin J. Minnich[b,c]* and Ahmed H. Zewail[a,b]†

[a]Physical Biology Center for Ultrafast Science and Technology, Arthur Amos Noyes Laboratory of Chemical Physics, California Institute of Technology, Pasadena, CA 91125
[b]Kavli Nanoscience Institute, California Institute of Technology, Pasadena, CA 91125
[c]Division of Engineering and Applied Science, California Institute of Technology, Pasadena, CA 91125

The dynamics of charge carriers in amorphous semiconductors fundamentally differ from those in crystalline semiconductors[1,2] due to the lack of long-range order and the high defect density. Despite intensive technology-driven research interests[3,4] and the existence of well-established experimental techniques, such as photoconductivity time-of-flight[5–8] and ultrafast optical measurements[9–12], many aspects of the dynamics of photo-excited charge carriers in amorphous semiconductors remain poorly understood. Here we demonstrate direct imaging of carrier dynamics in space and time after photo-excitation in hydrogenated amorphous silicon (a-Si:H) by scanning ultrafast electron microscopy (SUEM)[13,14]. We observe an unexpected regime of fast diffusion immediately after photoexcitation along with spontaneous electron-hole separation[15] and charge trapping[1] induced by the atomic disorder. Our findings demonstrate the rich dynamics of hot carrier transport in amorphous semiconductors that can be revealed by direct imaging based on SUEM.

---

* To whom correspondence should be addressed. Email: aminnich@caltech.edu
† Deceased

Charge carrier dynamics in amorphous semiconductors has been a topic of sustained research, particularly propelled by modern applications in thin film solar cells[16], transistors and optical sensors[4]. In amorphous semiconductors, the absence of long-range order leads to the breakdown of familiar concepts in crystalline materials related to the charge transport, such as Bloch states, reciprocal space and the momentum selection rule[1]. Instead, even in a fully coordinated and defect free amorphous semiconductor, Anderson localization[17] gives rise to localized electronic states near the edges of the conduction and the valence bands with a density of state decays exponentially with energy into the band gap[1]. These localized band-tail states are separated from extended states inside the bands at specific energies, called "mobility edges"[18]. In addition, in real amorphous semiconductors the high level of defects, such as dangling bonds and voids, contribute to the formation of deep defect states within the band gap[18]. Both band-tail and deep defect states can trap charge carriers in amorphous semiconductors, thus limiting the carrier mobility and lifetime of these materials[19]. Given the paramount importance of these "trap states" in determining the electrical properties of amorphous semiconductors, a complete understanding of the interactions between carriers and trap states has been a central pursuit in the field.

Amongst all amorphous semiconductors, hydrogenated amorphous silicon (a-Si:H) has served as an archetypical example due to its elemental simplicity and technological relevance[3,19,20]. Atomic hydrogen in a-Si:H passivates the silicon dangling bonds and largely reduces the density of the in-gap deep defect states, leading to significant improvement in electrical properties[19]. Numerous studies have been directed towards understanding the charge carrier dynamics in a-Si:H. Of particular interest are the dynamics of photo-excited carriers in a-Si:H, which directly affect the performance of a-Si:H-based thin film solar cells[16] and optoelectronic devices[4]. One widely used technique is the photoconductivity time-of-flight measurement in the nanosecond[5,6] to picosecond regimes[7,8]. In these measurements, the transient signal resulted from the photo-generated charge carriers inside an a-Si:H p-i-n junction device is recorded either electrically or optically, and the drift mobility of the charge carriers can be estimated using measured transit time, sample thickness, and applied electrical collection field. Alternatively, picosecond optical pump-probe measurements[9–11] have been used to characterize the recombination and trapping dynamics of charge carriers after photo-generation. However, these techniques are indirect as they infer the dynamics from secondary effects such as photo-induced absorption[9] or photo-bleaching of electroabsorption[7], complicating the interpretation of results and impeding efforts to directly trace the actual transport processes.

In this letter we report direct imaging of the charge carrier dynamics in a-Si:H after photo-excitation with scanning ultrafast electron microscopy (SUEM)[13,14,21,22]. SUEM is a pump-probe microscopy technique that combines the spatial resolution of the electron probe with the temporal resolution of the ultrafast laser[23]. SUEM is uniquely suited for studying charge carriers' spatiotemporal dynamics at the surfaces and interfaces of semiconductors. Previously, SUEM was used to image the entire process of charge carrier generation, transport and recombination at the silicon p-n junction, providing new insights into the ballistic transport across the junction[22]. In the present work, we observe striking evidence of an anomalously fast diffusion, compared to the rate expected from the bulk mobility, after photoexcitation, as well as spontaneous electron-hole separation[24] and a transition from diffusion to trapping for both electrons and holes. These observations are reproduced by numerical Monte Carlo simulations incorporating scattering and trapping events. Our study demonstrates the rich and unexpected dynamics of hot carrier transport in amorphous semiconductors that can be revealed by direct imaging based on SUEM.

Details of the operation and image interpretation of SUEM can be found in Methods. Figure 1 displays the SUEM images taken at different time delays between the laser pump and the electron probe, where the bright and the dark contrast indicate excess electron and hole populations, respectively (see Methods). For visual clarity, a low-pass Gaussian filter was used to remove high-spatial-frequency noise, while the raw images were used for the quantitative analysis elsewhere in the paper. In Fig. 1, we observe a bright contrast just after the arrival of the pump pulse, which after 20 ps transforms into a bright disk due to electron-hole-pair generation. Between 20 ps and 100 ps, the bright disk expands rapidly, while its center becomes dark. After 100 ps, the size of the resulting ring stabilizes, while its center continues to get progressively darker, peaking at 870 ps. From these images we identify two distinct regimes. At times shorter than 100 ps the ring forms and quickly expands, while after 100 ps the size of the ring stabilizes and the dark contrast at the center becomes increasingly prominent.

To quantitatively interpret this observation, we construct a simple model to describe the dynamics of electrons and holes after photo-excitation. As depicted in Fig. 2(a), we assume the electron and hole densities evolve as Gaussian distributions in space following the Gaussian profile of the laser beam intensity. Given the higher mobility of electrons[19], they transport out more rapidly than holes, giving rise to a net charge distribution with excess holes in the center and excess electrons in the surrounding area. Because excess electrons (holes) produce bright (dark) contrasts in SUEM, this net charge distribution leads to the observed ring shape. To compare the model with our observation, we extracted the spatial profiles along a stripe-shaped area in the SUEM images, as marked by the yellow lines in Fig. 2(b), and plotted as blue solid lines in Fig. 2(c). The subtraction of the Gaussian fits for electron and hole populations, representing the net charge density, are plotted as orange solid lines and show reasonable agreements with the experimental data.

The apparent charge imbalance between electrons and holes, evidenced by the formation of bright ring and dark center regions, indicates the absence of ambipolar diffusion. In most crystalline semiconductors, photo-excited electron-hole pairs tend to diffuse together with an intermediate diffusivity due to the Coulombic interaction between them. Instead, here we see spatial separation of electrons and holes after photo-excitation. The charge separation and the emergence of spatial distribution of net charges were predicted to happen in so-called "relaxation semiconductors"[15], including most wide-gap crystalline semiconductors and amorphous semiconductors. In these materials, the resistivity is usually so high that the dielectric relaxation time $\tau_d = (\varepsilon\varepsilon_0)/\sigma$ (here $\varepsilon$ is the relative permittivity, $\varepsilon_0$ is the vacuum permittivity and $\sigma$ is the electrical conductivity) can be significantly longer than the recombination time of the photo-excited carriers. The dielectric relaxation time determines the time scale over which a net charge distribution can be neutralized. In the case of a long dielectric relaxation time, the effect of the electric field resulting from charge separation is effectively weak and the local charge neutrality cannot be maintained. In particular, a case study of this phenomenon in a-Si:H was conducted by Ritter et al.[24].

From the fitted model we can evaluate the transport properties of electrons and holes. In Fig. 3(a) we plot the squared $1/e$ radius of the Gaussian distributions $l^2$ as a function of time $t$. In a normal (Gaussian) diffusion process, $l^2 = 4Dt$, where $D$ is the diffusivity. In Fig. 3a we give linear fits to the data below 100 ps. The deviation of experimental results from the linear fit suggests a "superdiffusive" behavior immediately after the photo-excitation. For comparison, we also plot quadratic fits in Fig. 3a, which agree better with the data. However, we caution that the exact time dependence here is not conclusive due to the limited signal-noise ratio of the experiment. We conducted another measurement with a higher fluence (see Supplementary Information), and observed the fast diffusion behavior as well with time dependence closer to linear. The linear fits indicate diffusivity values on the order of $10^3$ cm$^2$/s for both electrons ($8 \times 10^3$ cm$^2$/s) and holes ($3.8 \times 10^3$ cm$^2$/s), several orders of magnitude higher than those extracted from steady-state measurements[25]. A similar effect has been recently observed in crystalline silicon with SUEM[26], explained by the fast initial expansion of hot electron and hole densities. Particularly in a-Si:H, the average initial temperature of photo-excited electrons and holes can be estimated to be $(E_{ph} - E_G)/k_B \approx 8000$ K, where $E_{ph} = 2.4$ eV is the energy of incident photons and $E_G = 1.73$ eV is the optical band gap of a-Si:H[19]. Our observation in a-Si:H suggests that this process is not sensitive to the long-range disorder and thus can occur in amorphous materials, an intriguing conclusion that is worth further theoretical and experimental investigations. To the best of our knowledge, this fast diffusion behavior was not observed in earlier picosecond photoconductivity time-of-flight experiments on a-Si:H[7], possibly because in these experiments the photo-carriers diffuse equally to the positive and negative electrodes, generating no net electrical effects. Although an electrical collection field is usually used in these experiments, this field is not effective in driving carrier transport in "relaxation semiconductors" as discussed above[15]. The observation of this fast diffusion process in an amorphous semiconductor is surprising and may have important implications with regard to optoelectronic device applications, since the performance of amorphous-semiconductor-based optoelectronic and photovoltaic devices is largely limited by the poor charge transfer ability of the amorphous semiconductors.

After the initial fast expansion within 100 ps, there is a clear transition of the dynamics of both electrons and holes: the widths of their distributions stop increasing and stabilize at times up to 2 ns, as shown in Fig. 3(a). We interpret this distinct behavior as the trapping of hot carriers as they cool down to the band-tail localized states and/or deep in-gap defect states, a well-known feature of amorphous semiconductors that limits the drift mobility of charge carriers[1]. In particular, the change of the intensities of the bright ring and dark center, as shown in Fig. 3(b), suggests details of the trapping process. The intensity within the ring, where hot electrons reside, decreases monotonically, indicative of the trapping of energetic electrons with a time scale of hundreds of picoseconds. The fit to the experimental data shown in Fig. 3(b) is cubic-polynomial; the data cannot be fitted with an exponential function satisfactorily, suggesting that the trapping process cannot be described by a single time constant. Simultaneously, the contrast of the dark center region first gets darker till 900 ps, and then slowly becomes less dark. This behavior implies that the hot electrons are trapped in a faster time scale than the hot holes, resulting in an initial increase in the dark contrast. Beyond 900 ps, the trapping of hot holes becomes appreciable and the dark contrast starts to reduce. The excitation and trapping processes are schematically shown in Fig. 3(c); in this experiment the trapping into the localized band-tail states and into the deep defect states is not resolved. In an ultrafast optical pump-probe measurement of a-Si:H, Vardeny et al.[9] observed exponential decay of photo-induced absorption in a similar time scale, which they interpreted as the trapping process of photo-excited carriers, in agreement with our observation; however, they were unable to identify the separate behaviors of electrons and holes. We note that the carrier recombination occurs over a longer time scale (a few nanoseconds) than the time window of our experiment at this carrier concentration for which Auger recombination is weak[11], and so little recombination is observed over the time delays measured here.

A quantitative transport model of the hot carrier expansion process in a-Si:H is currently not accessible due to the lack of a clear understanding of charge scattering mechanisms and a practical formalism of electronic transport in amorphous semiconductors. In particular, the Boltzmann transport equation, which is the standard tool used in crystalline materials, is known to be inapplicable in amorphous semiconductors[2]. Instead, we attempt to unify the observed fast diffusion and trapping processes using a phenomenological Monte Carlo simulation. See Methods for details of the simulation. Briefly, we initiate the expansions of the hot electron and hole gases at a starting temperature of 8000 K. The kinetic energies of electrons and holes are damped at constant rates, representing inelastic scattering events that cool down the electron/hole gases. Simultaneously, the travel directions of the particles are randomized at each time step with a certain probability set by a characteristic lifetime associated with elastic scattering events. The randomized travel directions also follow a probability distribution that favors small-angle scatterings, typical of point defect scattering[27]. Whenever the kinetic energy of a specific particle drops below a certain threshold ("mobility edge"), this particle is deemed "trapped" and fixed in space. In Fig. 3(a), the solid curves represent the second moments $\langle r^2 \rangle$ of the electron and hole distributions as a function of time from the Monte Carlo simulation. We also show in Fig. 4 the simulated SUEM images at different time delays (Fig. 4(a) to (c)), in qualitative agreement with our experimental results, as well as the time evolution of the radial distribution functions of electrons and holes in the simulation (Fig. 4(d) and (e)), which clearly shows the slowdown of the initial fast expansion process and the transition into the trapping dynamics.

In conclusion, we have directly imaged the dynamics of photo-excited hot carriers in a-Si:H at ultrafast timescales by SUEM. We observe an unexpected regime of fast diffusion immediately after photoexcitation, likely due to the initial high temperature of photoexcited carriers, followed by trapping of both electrons and holes. Our observations are in good qualitative agreement with a transport model based on phenomenological Monte Carlo simulations. Furthermore, to the best of our knowledge, our observation of the spontaneous electron-hole separation is the first direct verification of the "relaxation semiconductor" behavior predicted in the 1970s[15]. This work demonstrates the power of the SUEM to provide new insights into hot carrier dynamics in diverse materials.

**Methods**

**Sample Preparation**

The sample studied is a-Si:H (thickness ~100 nm) grown by PECVD on a crystalline silicon substrate with a 1-µm-thick thermal oxide layer. The sample is further characterized by Raman spectroscopy (see Supplementary Information).

**Scanning Ultrafast Electron Microscopy**

The details of SUEM setup have been reported elsewhere[13,14]. Briefly, infrared laser pulses (1030 nm, 300 fs), generated by a Clark-MXR fiber laser system, are split to generate green (515 nm) and UV (257 nm) pulses; the green is focused onto the sample as both the photo-excitation and the clocking pulse, while the UV pulse is focused onto the photocathode to generate ultrashort electron pulses. The diameter of the pump is ~60 µm, and the fluence at 515 nm (photon energy 2.4 eV) for data reported in the main text is ~20 µJ/cm$^2$ with the pulse repetition rate at 25 MHz, corresponding to a peak carrier concentration $\sim 3\times 10^{18}$ cm$^{-3}$ [28]. The data for a fluence of ~67 µJ/cm$^2$ (repetition at 5 MHz) is reported in the Supplementary Information.

The electron pulses, which are accelerated at 30 kV, are delayed by a delay stage from -680 ps to 3.32 ns after the photo-excitation pulses and are spatially rastered over the region of interest to form an image. The electron pulses incident on the sample produces secondary electrons (SEs) from the top 1-10 nm of the sample, which are collected by a positively biased Everhart-Thornley detector. To enhance the signal, the background is removed by subtracting a reference image recorded prior to optical excitation (at -680 ps), from the images recorded at different time delays. This results in so-called "contrast images", in which bright and dark contrasts are interpreted as increased electron and hole concentrations, respectively. For a given material, the number of emitted SEs also depends on the surface topography, chemical composition, and local fields. The removal of the background ensures that the observed contrast reflects only the changes in local carrier density due to the optical excitation.

**Monte Carlo Simulation**

$10^6$ electrons and holes are included in the simulation, which are first randomly assigned the positions and velocities from Gaussian distributions determined by the beam radius and temperature, respectively. The effective masses used for electrons and holes are $0.3m_0$ and $m_0$, respectively[19]. Subsequently the motion of each particle is tracked in the simulation. The kinetic energy of each particle is damped at a constant rate (28 ps for electrons and 53 ps for holes). The travel direction of each particle is also randomized at each time step with a probability of $\Delta t/\tau_{\text{elastic}}$, where $\Delta t$ is the length of the time step (100 fs in the simulation), and $\tau_{\text{elastic}}$ is a characteristic lifetime associated with elastic scattering events (0.8 ps for electrons and 1 ps for holes). The scattering angle follows a Gaussian distribution with a width of 45 degrees to favor small-angle scatterings. When the kinetic energy of a particle drops below the mobility edge (0.1 eV for electrons and 0.25 eV for holes), the particle is "trapped" and fixed in space. We emphasize here that the parameters used here are purely phenomenological and are chosen to give the best fits to experimental results, as shown in Fig. 3(a). The second moments of the distributions of electrons and holes $\langle r^2 \rangle$ are calculated using the following formula

$$\langle r^2 \rangle = \frac{\int r^2 f(r) d^2\mathbf{r}}{\int f(r) d^2\mathbf{r}} = \frac{\int_0^{+\infty} r^2 \phi(r) dr}{\int_0^{+\infty} \phi(r) dr}, \tag{1}$$

where **r** is the position vector, $f(r)$ is the number of electrons per unit area, and $\phi(r) = 2\pi r f(r) \Delta r$ is the radial distribution function, which counts number of particles within a differential ring region (a "bin") with width $\Delta r$ and radius $r$. The radial distribution functions for electrons and holes are plotted in Fig. 4(d) and Fig. 4(e), respectively, and they intuitively show the slowdown of the initial fast diffusion and the transition into the trapping regime.

## Acknowledgements

We thank Yangying Zhu for providing the sample, and Xuewen Fu for helpful discussions. This work is supported by the National Science Foundation (DMR-0964886) and the Air Force Office of Scientific Research (FA9550-11-1-0055) in the Gordon and Betty Moore Center for Physical Biology at the California Institute of Technology. B. L. is grateful to the financial support from the KNI Prize Postdoctoral Fellowship in Nanoscience at the Kavli Nanoscience Institute of California Institute of Technology.

## Author Contributions

B. L., E. N. and H. L. did the experiment and analyzed the results. B. L. wrote the paper. A. J. M. proofread, commented on the manuscript and advised on the modeling work. A. H. Z. supervised the research.

## Competing Financial Interests

The authors declare no competing financial interests.

**Figures and Captions**

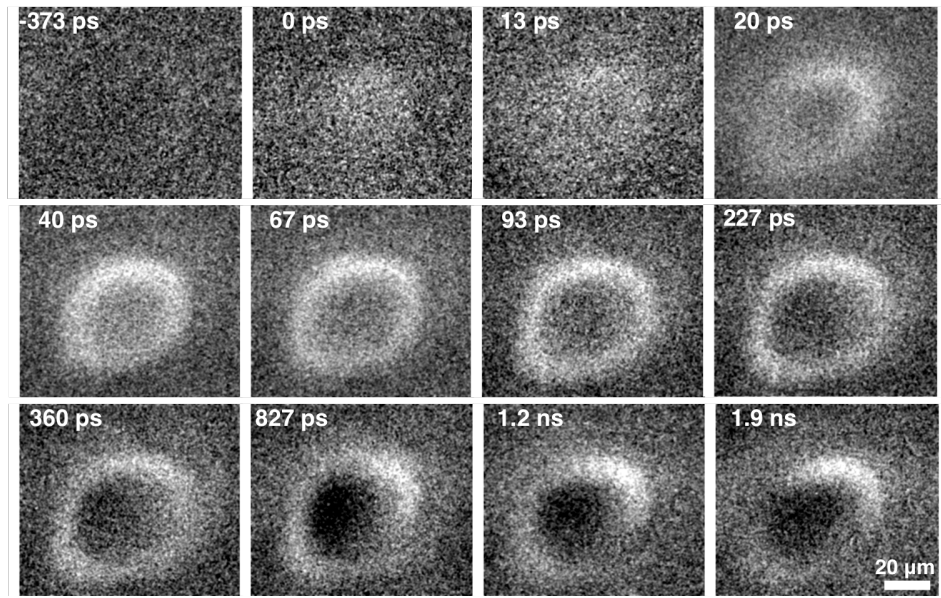

**Figure 1 SUEM images at different delay times after the photo-excitation.** Images shown here are "difference images" with an image at -730ps as the reference. Each image represents an average of 60 to 120 images at the same delay time. Raw images are filtered with a low-pass Gaussian filter to suppress high-spatial-frequency noise and enhance the visual contrast.

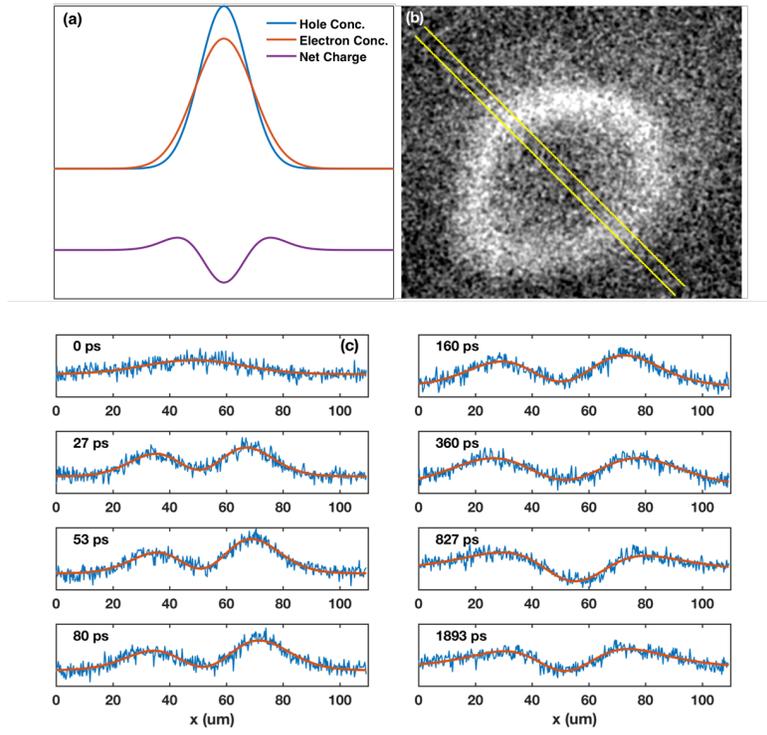

**Figure 2 Analysis of the image intensity along a center-cut line.** (a) An illustration of the model used to interpret the experimental observation. The blue and orange lines are the spatial distributions of hole and electron concentration, respectively. The purple line is the difference of the hole and electron distributions, namely the net charge distribution. (b) The yellow markers indicate the region within which the line-cuts are selected and averaged. (c) The averaged intensity distribution within the strip region shown in (b): the blue lines are the experimental data, while the orange line is the least-square fit with the model in (a).

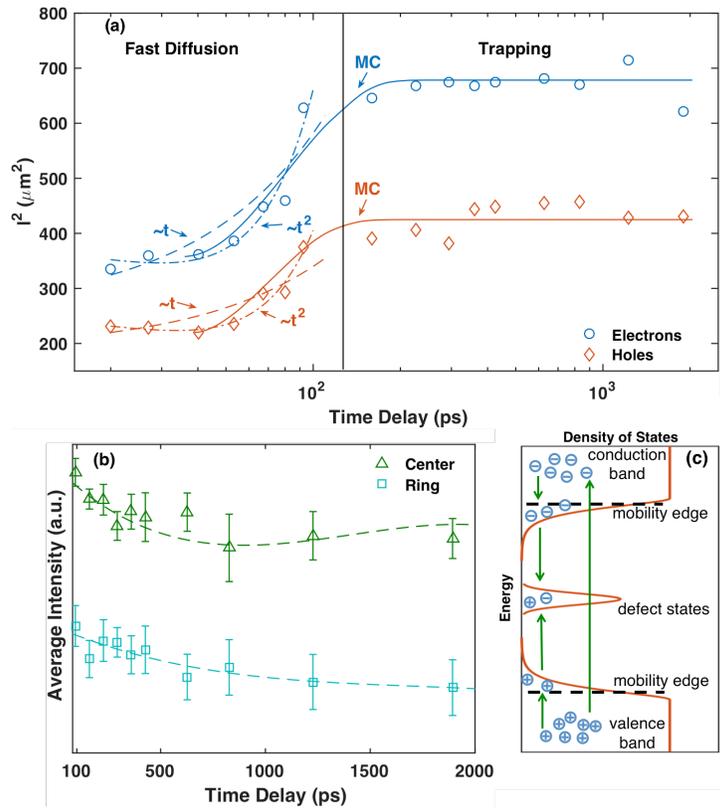

**Figure 3 Quantitative analysis of carrier diffusion and trapping processes.** (a) The squared $1/e$ radius $l^2$ of the spatial distributions of electrons and holes versus the delay time. The dashed/dot-dashed lines are linear/quadratic fits to the experimental data before 100 ps. The solid lines (labeled "MC") are results of the Monte Carlo simulation for the entire dynamic process. (b) The average intensities of the bright ring region and the dark central region versus the time delay after 100 ps. The dashed lines are cubic-polynomial fits to guide the eye. The error bars represent the standard deviation of the intensity distribution within corresponding areas. (c) An illustration of the typical density of states and the excitation and trapping processes (green arrows) in a-Si:H. The dark dashed lines mark the mobility edges separating the localized band-tail states and extended states in both conduction and valance bands. The blue circles with "+" and "-" signs represent holes and electrons, respectively.

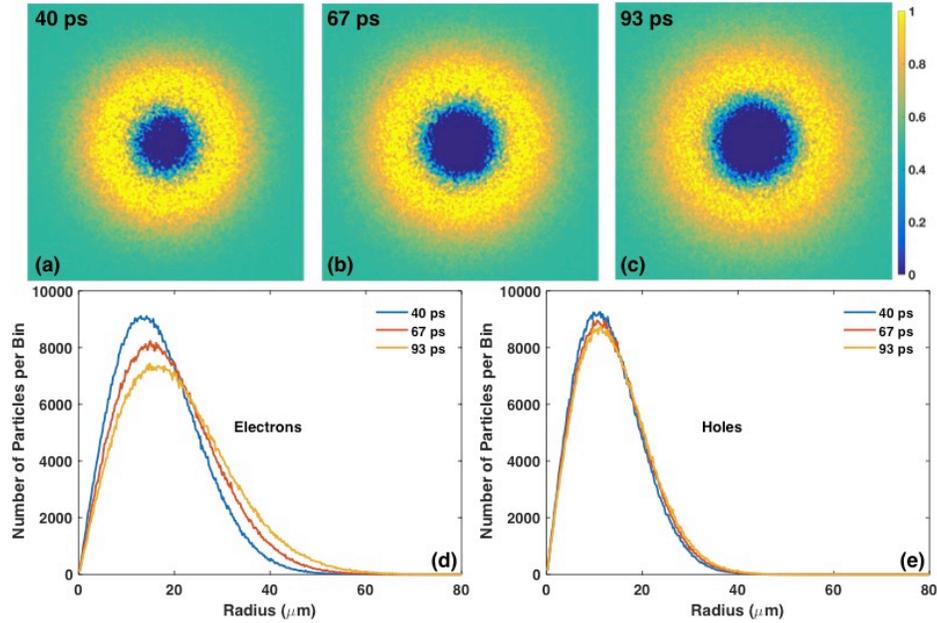

**Figure 4 Monte Carlo simulation of the carrier dynamics.** The simulated SUEM images at the time delays of (a) 40 ps, (b) 67 ps and (c) 93 ps. The intensity of the image represents net distributions of electrons (above 0.5 on the color bar) and holes (below 0.5 on the color bar). (d) and (e) are the radial distribution functions of electrons and holes at different time delays. The radial distribution functions are defined in Methods, and count the number of particles within differential rings ("bins") with a width of 200 nm and a radius corresponding to the horizontal axes. The radial distribution functions intuitively show the slowdown of the initial fast diffusion and the transition into trapping.

# Supplementary Information

# Dynamics of Photo-excited Hot Carriers in Hydrogenated Amorphous Silicon Imaged by 4D Electron Microscopy

Bolin Liao[a,b], Ebrahim Najafi[a], Heng Li[a], Austin J. Minnich[b,c,‡] and Ahmed H. Zewail[a,b,§]

[a]Physical Biology Center for Ultrafast Science and Technology, Arthur Amos Noyes Laboratory of Chemical Physics, California Institute of Technology, Pasadena, CA 91125
[b]Kavli Nanoscience Institute, California Institute of Technology, Pasadena, CA 91125
[c]Division of Engineering and Applied Science, California Institute of Technology, Pasadena, CA 91125

---

[‡] To whom correspondence should be addressed. Email: aminnich@caltech.edu
[§] Deceased

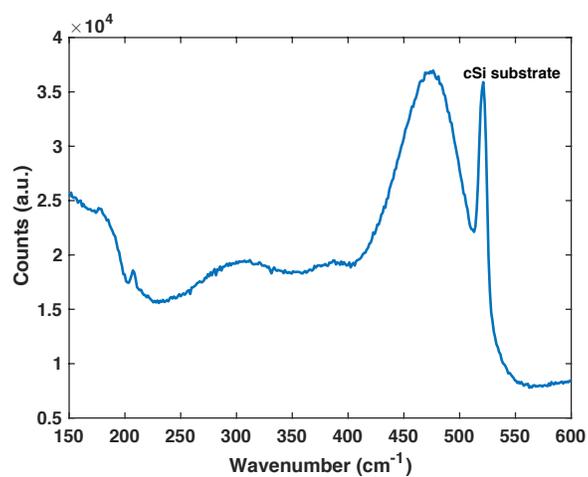

**Supplementary Figure 1. Raman Characterization of the Sample.** The Raman spectrum of the sample shows the wide "optical peak" of the amorphous silicon thin film at 475 cm$^{-1}$ (see Beeman et al., *Physical Review B*, **32**, 874 (1985)) and the narrow peak of the crystalline silicon substrate at 520 cm$^{-1}$.

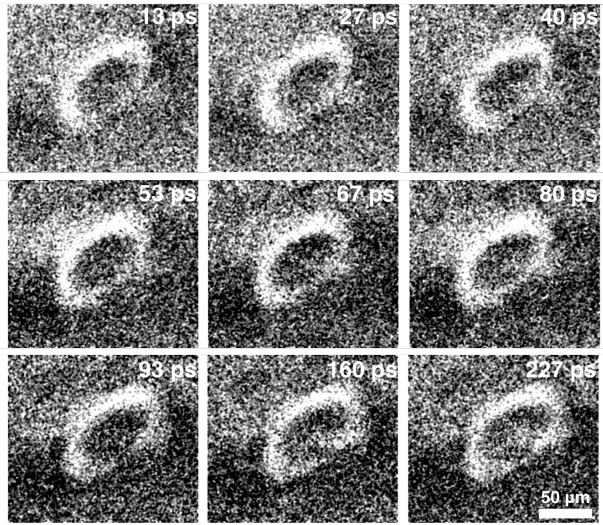

**Supplementary Figure 2. SUEM Images Taken at a Higher Pump Fluence.** The data is measured with a pump fluence of ~67 μJ/cm$^2$ (repetition rate at 5 MHz). The feature is significantly larger than that with the lower fluence as reported in the main text. This is because higher density of electrons and holes are excited above the detection threshold of the SUEM.

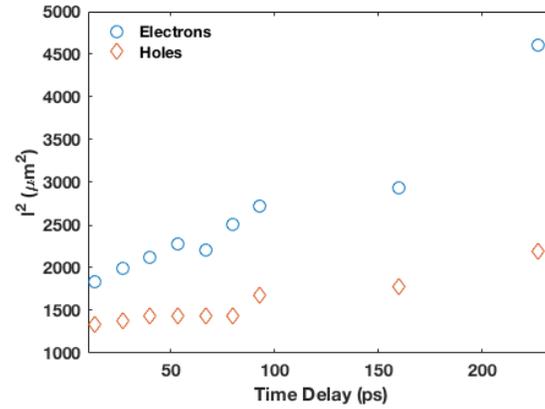

**Supplementary Figure 3. Fast diffusion of electrons and holes at the higher pump fluence.** The squared radii of the electron and hole distributions with the higher pump fluence (67 µJ/cm$^2$) are shown as a function of time delay. In comparison to the data at the lower fluence reported in the main text, the time dependence here is closer to be linear, than quadratic, whereas the average speed of expansion is similar.